\documentclass[twocolumn,prl,showpacs,preprintnumbers,amsmath,amssymb,superscriptaddress]{revtex4-1}
\usepackage{graphicx}
\usepackage{setspace}

\usepackage{color}
\usepackage{ulem} 

\usepackage{amsmath,amsthm,amssymb}
\usepackage{mathrsfs}

\newcommand{\udarrow}{\uparrow\downarrow}

\makeatletter
\def\Left#1#2\Right{\begingroup%
   \def\ts@r{\nulldelimiterspace=0pt \mathsurround=0pt}%
   \let\@hat=#1%
   \def\sht@im{#2}%
   \def\@t{{\mathchoice{\def\@fen{\displaystyle}\k@fel}%
          {\def\@fen{\textstyle}\k@fel}%
          {\def\@fen{\scriptstyle}\k@fel}%
          {\def\@fen{\scriptscriptstyle}\k@fel}}}%
   \def\g@rin{\ts@r\left\@hat\vphantom{\sht@im}\right.}%
   \def\k@fel{\setbox0=\hbox{$\@fen\g@rin$}\hbox{%
      $\@fen \kern.3875\wd0 \copy0 \kern-.3875\wd0%
      \llap{\copy0}\kern.3875\wd0$}}%
      \def\pt@h{\mathopen\@t}\pt@h\sht@im%
      \Right}%
\def\Right#1{\let\@hat=#1%
   \def\st@m{\mathclose\@t}%
   \st@m\endgroup}
\makeatother

\begin{document}
\title{High-Harmonic Generation in Mott Insulators}

\author{Yuta Murakami}
\affiliation{Department of Physics, University of Fribourg, 1700 Fribourg, Switzerland}
\author{Martin Eckstein}
\affiliation{Department of Physics, University of Erlangen-N\"urnberg, 91058 Erlangen, Germany}
\author{Philipp Werner}
\affiliation{Department of Physics, University of Fribourg, 1700 Fribourg, Switzerland}
\date{\today}

\begin{abstract}
Using Floquet dynamical mean-field theory, we study the high-harmonic generation in the time-periodic steady states of wide-gap Mott insulators under AC driving. 
In the strong-field regime, the harmonic intensity exhibits 
multiple plateaus, whose cutoff 
energies 
$\epsilon_{\rm cut} = U + mE_0$ 
scale with the Coulomb interaction $U$ and the maximum field strength $E_0$.
In this regime, the created doublons and holons are localized because of the strong field and the 
$m$-th plateau originates 
from the recombination of $m$-th nearest-neighbor doublon-holon pairs.
In the weak-field regime, there is only a single plateau in the intensity, which originates from the recombination of itinerant doublons and holons. 
Here, $\epsilon_{\rm cut} = \Delta_{\rm gap} + \alpha E_0$, with $\Delta_{\rm gap}$ the band gap and $\alpha>1$. 
We demonstrate that the Mott insulator shows a stronger high-harmonic intensity than a semiconductor model with the same dispersion as the Mott insulator,
even if the semiconductor bands are broadened by impurity scattering to mimic the incoherent scattering in the Mott insulator.
\end{abstract}


\maketitle
{\it Introduction--}
The interaction between intense laser fields and matter results in highly nonperturbative phenomena.
Among them, the high-harmonic generation (HHG) is 
both interesting with regard to the underlying physical processes and in view of potential applications~\cite{Corkum1993,Lewenstein1994,Cavalieri2007,Krausz2009}.
HHG in atomic and molecular gases has been intensively studied for decades,
and is the basis of attosecond science and new ultrafast imaging methods~\cite{Cavalieri2007,Krausz2009}.
The recent observation of HHG in semiconductors has renewed the interest in this field~\cite{Ghimire2010,Schubert2014,Hohenleutner2015,Luu2015,Vampa2015b,Langer2016,Ndabashimiye2016,Liu2016,You2016,Yoshikawa2017}.
Originating from the periodic arrangement of the atoms in solids, characteristic features of the HHG spectra, different from those of gases, have been observed.
HHG in semiconductors can be used to explore the electron band properties \cite{Ghimire2010,Ndabashimiye2016,Liu2016,Hohenleutner2015} and the lattice structure~\cite{You2016},
to probe electron dynamics on ultrafast time scales~\cite{Vampa2015b,Ndabashimiye2016}, 
and to develop new high-frequency laser light sources \cite{Ndabashimiye2016}. 
Theoretically, several mechanisms for HHG in solids have been proposed assuming weak correlations or an effective single-particle picture  ~\cite{Golde2008,Ghimire2010,Schubert2014,Hohenleutner2015,Luu2015,Kemper2013b,Higuchi2014,Vampa2014,Vampa2015,Vampa2015b,Langer2016,Ndabashimiye2016,Liu2016,You2016,Tamaya2016,Otobe2016,Luu2016,Yoshikawa2017,Ikemachi2017,Tancogne-Dejean2017b,Tancogne-Dejean2017},
such as intraband electron dynamics, interband contributions from electron-hole recombination~\cite{Vampa2014,Vampa2015,Vampa2015b}, and time-dependent diabatic processes~\cite{Tamaya2016,Yoshikawa2017}. 

A different class of insulators in solid state physics
 is the Mott insulator (MI), which originates from strong electronic correlations,
 and the possibility of HHG in MIs has recently been pointed out \cite{Ivanov2018,Tancogne-Dejean2017c}.  
In MIs, the excitation creates doublons and holons instead of electrons and holes in semiconductors, and their dynamics determines the current and the HHG.
However, in contrast to semiconductors, excited charges cannot move freely in MIs because 
of Pauli blocking and scattering. 
Therefore, the features of the high-harmonic spectrum of MIs are not a priori clear, and the current understanding of HHG in MIs is very limited. Deeper insights into the underlying physics 
may lead to 
applications of HHG in the ultrafast 
imaging of the carrier dynamics in strongly correlated systems \cite{Ivanov2018} and open a new class of materials for use in light sources. 

In this work, we shed light on the periodically driven MI phase of the half-filled single-band Hubbard model.  
By means of the nonequilibrium dynamical mean-field theory (DMFT) \cite{Aoki2013}, we reveal the general and fundamental structure of the HHG spectrum and its relation to the dynamics of the doublons and holons.
Moreover, by comparing the HHG in MIs and semiconductor models, we find a different relation between the single particle spectrum and the HHG spectrum in these insulators.

{\it Formalism--}
 \begin{figure*}
  \centering
    \hspace{-0.cm}
    \vspace{0.0cm}
   \includegraphics[width=165mm]{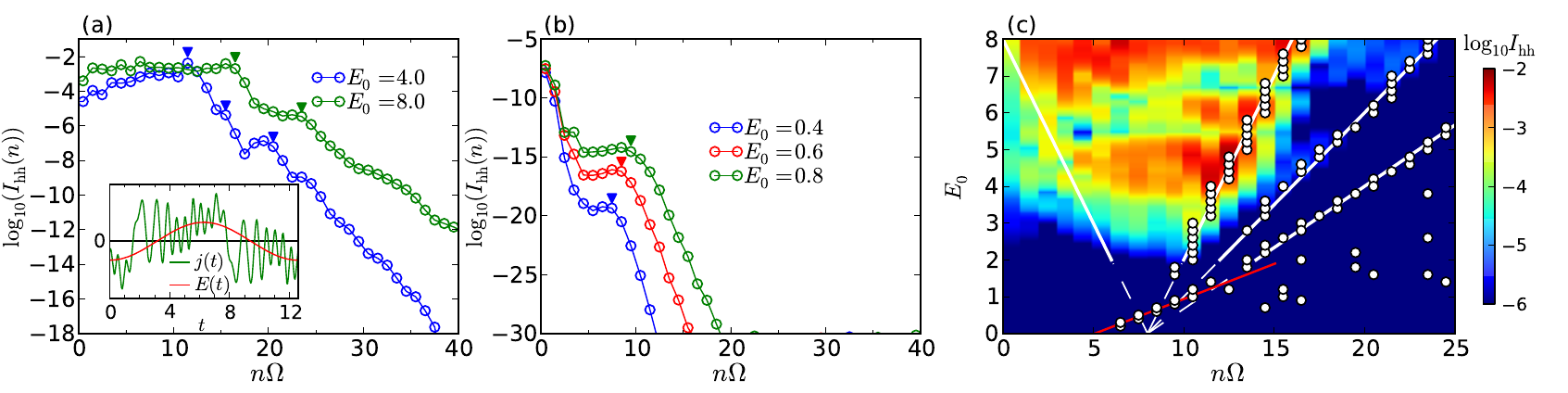} 
  \caption{(a) HHG spectra in the strong-field regime, (b) HHG spectra in the weak-field regime, and (c) HHG spectra as a function of the field strength ($E_0$) and the harmonic energy ($n\Omega$). The arrows and the white circle markers show cutoff energies. In panels (a)(b), we also show the contribution from $j_{\rm rc}$, which is evaluated by the generalized tunneling formula. White lines in panel (c) indicate $n\Omega=U+mE_0$ and the red line is the fit for the weak-field regime. The inset of panel (a) shows the current and electric field during one period for $E_0=4.0$. We use $U=8.0,\beta=2.0,\Gamma=0.06,\Omega=0.5$.
  }
  \label{fig:fig1}
\end{figure*}
We consider the Hubbard model attached to a 
thermal bath and driven by an AC field,
\begin{align}
H=&-\sum_{\langle i,j\rangle,\sigma} v_{ij}(t) c_{i,\sigma}^\dagger c_{j,\sigma}+U\sum_i n_{i\uparrow}n_{i\downarrow}+H_{\rm bath}.
\end{align}
Here $c^\dagger_{i,\sigma}$ is the creation operator of an electron at site $i$ with spin $\sigma$, $v_{ij}$ indicates the hopping parameter, $U$ is the interaction and $q$ is the charge.
In the calculations, we use the gauge with pure vector potential ${\bf A}(t)$ so that
the effect of the electric field ${\bf E}(t)$ appears in the phase of the hopping parameter 
$v_{ij}(t)=v_{ij}\exp\bigl(-iq\int^{{\bf r}_j}_{{\bf r}_i} d{\bf r} {\bf A}(t)\bigl)$, where ${\bf A}(t)$ is related to the electric field by ${\bf E}(t)=-\partial_t {\bf A}(t)$. 
This is equivalent to a  pure scalar potential term $\sum_{i,\sigma}\Phi({\bf r}_i,t) c^\dagger_{i\sigma}c_{i\sigma}=-{\bf E}(t)\cdot (\sum_{i,\sigma}{\bf r}_i c^\dagger_{i\sigma}c_{i\sigma})$
in the Hamiltonian.
$H_{\rm bath}$ represents a thermal bath of noninteracting electrons (the B\"{u}ttiker model), which is introduced to describe the coupling of the system to an environment~\cite{Tsuji2008,Mikami2016,Murakami2017}.
When the system is continuously excited by an external field with frequency $\Omega$,
it reaches a time-periodic nonequilibrium steady state (NESS) with a period $\mathcal{T}\equiv\frac{2\pi}{\Omega}$,
when the energy injected by the field is balanced by the dissipation to the bath.

We consider a hyper-cubic lattice with lattice spacing $a$ in the limit of infinite spatial dimensions ($v=\frac{v^*}{2\sqrt{d}}$ with $d\rightarrow\infty$), 
which has a Gaussian density of states $\rho(\epsilon)=\frac{1}{\sqrt{\pi}v^*}\exp[-\epsilon^2/v^{*2}]$.
The field is applied along the body diagonal, ${\bf A}(t)=A(t){\bf e}_0$ with ${\bf e}_0=(1,1,\cdots,1)$ and $qaA(t)=A_0\sin \Omega t$, 
so that its strength along a given axis is $E(t)=-\frac{A_0}{qa}\Omega \cos\Omega\equiv -E_0\cos\Omega t$.
For the thermal bath, we employ the B\"{u}ttiker model
with a finite band width $W_{\rm bath}$,
$-{\rm Im}\Sigma^R_{\rm bath}(\omega)=\Gamma \sqrt{1-\left(\omega/W_{\rm bath}\right)^2}$.
In the following we set $q,a=1$ and use $v^*$ as the unit of energy. 
In order to clarify fundamental aspects of HHG in MIs, we focus on systems where the Mott gap is large compared to the width of the Hubbard bands, 
and the excitation frequency is much smaller than the gap. 
We typically use $U=8,\beta=2.0,\Gamma=0.06,W_{\rm bath}=5,\Omega=0.5$.

To analyze the HHG spectrum, we focus on NESS calculated within Floquet dynamical mean-field theory (FDMFT) \cite{Schmidt2002,Joura2008,Tsuji2008,Tsuji2009,Lee2014,Mikami2016,Murakami2017,Max2017}\cite{Note2}. 
We implement the FDMFT method with the non-crossing approximation (NCA) as an impurity solver \cite{Eckstein2010b,supplemental}.
NCA is the lowest order self-consistent hybridization expansion and is expected to produce qualitatively correct results for large $U$.

{\it Results--}
 In Fig.~\ref{fig:fig1}(a)(b), we show the HHG spectra in the strong field regime ($E_0\gtrsim 2$) and the weak field regime ($E_0\lesssim 1$).
 The HHG spectrum is evaluated from the square of the Fourier transformation of the dipole acceleration $\frac{d}{dt}j(t)$
as $I_{\rm hh}(n\Omega)=|n\Omega j(n\Omega)|^2$~\cite{Kemper2013b,Tamaya2016,Ivanov2018} with  $n\in \mathbb{Z}$,
which is proportional to the power radiated at the given frequency. 
Here, the current is defined as $j(t)=iq \sum_{i,j,\sigma} v_{ij}(t)({\bf e}_0 \cdot {\bf r}_{i-j}) \langle c_{i,\sigma}^\dagger(t) c_{j,\sigma}(t)\rangle
={\bf e}_0\cdot {\bf j}(t)$ and $j(n\Omega)=\frac{1}{\mathcal{T}}\int^{\mathcal{T}}_0 d\bar{t} e^{i\bar{t}n\Omega} j(\bar{t})$. 
In the inset of Fig.~\ref{fig:fig1}, we show an example of the time evolution of the electric field and the induced current during one period.
 Because of the inversion symmetry, only odd frequency components appear in the HHG spectrum. 

When the field is strong, the HHG spectrum 
initially increases with increasing order $n$
and exhibits a wide plateau, see Fig.~\ref{fig:fig1}(a).
 After this first plateau, the intensity suddenly drops, but other plateau structures exist at higher harmonic energies.
 On the other hand,  when the field is weak, the HHG spectrum first drops and then shows a plateau, after which the intensity vanishes exponentially, see Fig.~\ref{fig:fig1}(b).
 In both regimes, the cutoff energies monotonically increase with increasing field strength.
 
In Fig.~\ref{fig:fig1}(c), we show the HHG spectra as a function of $E_0$ and the harmonic energy ($n\Omega$). 
The cutoff energies of the plateaus are indicated by white markers \cite{Note1}.
The HHG spectra have nontrivial structures: 
i) the intensity  is strong in the triangular region $U-E_0\lesssim n\Omega\lesssim U+E_0$,
 ii) there is an enhanced intensity around $E_0=U/2=4$,
and 
 iii) the intensity is suppressed for $5\lesssim E_0\lesssim 6$.
In the strong-field regime, the cutoff energy scales as $\epsilon_{\rm cut,m}=U+mE_0$.
On the other hand, in the weaker field regime the cutoff energy of the first plateau $\epsilon_{\rm cut,1}$ scales as $\epsilon_{\rm cut,1}=\Delta+\alpha E_0$, 
where $\Delta$ is an offset with $\Delta\neq U$ and $\alpha>1$ is not integer. 
These features are generic, as we confirmed by changing $\Omega$ and $U$ \cite{supplemental}.

We now discuss the origin of the HHG in MIs.
There are two contributions to the current: the doublon/holon hopping ($j_{\rm hop}$) and the doublon-holon recombination/creation ($j_{\rm rc}$)~\cite{supplemental}.
The former is analogous to the intraband current in a semiconductor, while the latter corresponds to the interband current, which represents recombination/creation of electron-hole pairs.
One can approximately evaluate both contributions by means of a generalized tunneling formula for NESSs \cite{Lee2014,supplemental,Murakami2018}, which works quantitatively very well in the parameter regime considered here. Such an analysis shows that  the contribution from the recombination of doublons and holons ($j_{\rm rc}$) dominates the current and is responsible for the plateaus both in the weak and strong field regimes~\cite{supplemental}.

\begin{figure}[t]
  \centering
    \hspace{-0.cm}
    \vspace{0.0cm}
   \includegraphics[width=86mm]{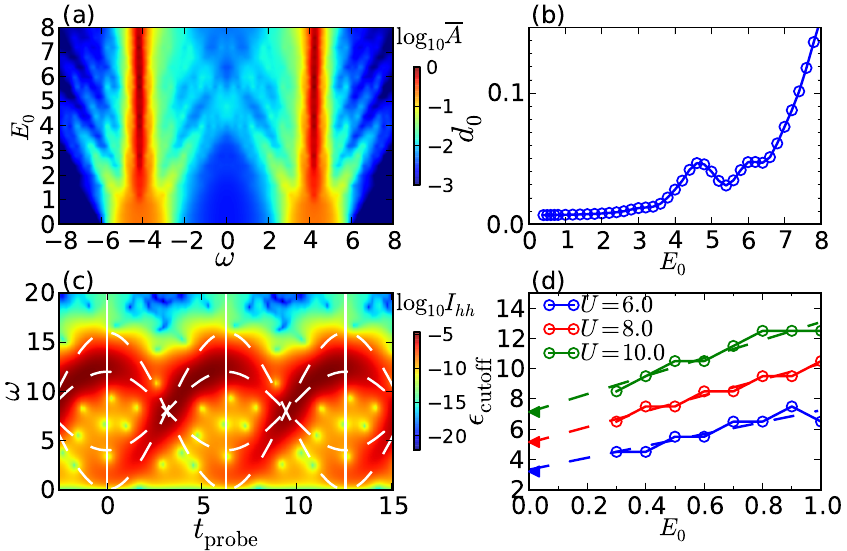} 
  \caption{(a) Time-averaged local spectral function ($\bar{A}(\omega)$) of the nonequilibrium steady state as a function of $E_0$. (b) Field-strength dependence of the time-averaged doublon density 
  ($d_0=\overline{\langle n_{\uparrow}n_{\downarrow}\rangle}$) in the NESS. (c) Log-scale plot of the temporal HHG intensity $I_{\rm hh}(\omega;t_{\rm probe})$ for $\Omega=0.5,E_0=4.0$. 
The dashed lines are $\omega=U\pm mE(t)$. Vertical lines indicate $t_{\rm probe}=0,\mathcal{T}/2,\mathcal{T}$.  
  Here $U=8.0,\beta=2.0,\Gamma=0.06$ and $W_{\rm bath}=5$.
  (d) Field strength dependence of the cutoff energy in the weak-field regime for various $U$. Dashed lines are linear fits and the arrows at $E_0=0$ indicate the gaps estimated from the local spectral functions.
  }
  \label{fig:fig2}
\end{figure}

In the strong field regime, this scenario of a dominant recombination/creation current is further supported by the fact that the cutoff energy is proportional to $U$ (the contribution from doublon/holon hopping should not depend on this energy scale \cite{Note3}).
We can thus argue that the different HHG plateaus originate from the recombination of a doublon-holon pair which is separated by $m$ sites: 
When $E_0$ is comparable or larger than the width of the Hubbard bands,
doublons and holons remain almost localized.
Indeed, the spectral functions in the NESS show clear Wannier-Stark peaks in the strong field regime~\cite{Lee2014,Werner2015}, see Fig.~\ref{fig:fig2}(a).
The recombination of a doublon-holon pair separated by $m$ sites along the positive field direction releases the energy $U+m|E(t)|$ at a certain time $t$.
 This scenario consistently explains the main characteristic features of the HHG spectra in the strong-field regime. 
 Since $E(t)$ oscillates between $-E_0$ and $E_0$, one expects that the energy emitted from the recombination of an $m$-th nearest-neighbor doublon-holon pair lies  in the triangler region
  $U-mE_0\leq n\Omega \leq U+mE_0$, which naturally explains the 
 prominent HHG in the dominant $m=1$ sector, and the weaker cutoffs at larger $m$.
 Secondly, when the doublon and holon density is small (large), there are less (more) doublons/holons to recombine, which leads to a low (high) HHG intensity. 
We indeed find that in the NESS the doublon/holon number is suppressed around $E_0\simeq 5.5$, see Fig.~\ref{fig:fig2}(b), which explains the valley in the HHG spectrum.
The decrease of the double occupancy in the energy range $U/2\lesssim E_0 \lesssim U$ comes from the absence of resonant tunneling processes. 
Similarly, the intense HHG spectrum around $E_0=U/2=4$ is explained by an enhanced number of doublons.
The peak in $I_{\rm hh}$ and in the time-averaged doublon number is slightly shifted upward relative to $E_0=U/2$. 
This can be explained 
by the oscillation of the field, which implies that larger field strengths are necessary for efficient tunneling to the next nearest neighbor site. 

The scenario of recombination of $m$-th neighbor doublon-holon pairs is also confirmed by investigating the temporal structure of the HHG signal.
We performed a windowed Fourier transformation of $j(t)$, 
$j(\omega;t_{\rm probe})=\int d\bar{t} e^{i\bar{t}\omega}j(\bar{t}) W(\bar{t};t_{\rm probe})$
and evaluated $I_{\rm hh}(\omega;t_{\rm probe})\equiv |\omega j(\omega;t_{\rm probe})|^2$.
Here $W(t;t_{\rm probe})$ is the Blackman window function with a half-window of length $2$ centered at $t=t_{\rm probe}$.
In Fig.~\ref{fig:fig2}(c), we show the result for $E_0=4.0$ on a logarithmic scale. 
The intensity peak at each $t_{\rm probe}$ essentially follows $\omega=U\pm E(t)$,
 and we observe a sudden drop of the intensity near $\omega=U\pm2E(t)$.
 In particular, the $\omega=U+E_0$ and $\omega=U+2E_0$ components are strong around $|E(t)|=E_0$, as expected from the scenario above. 
 
In the weak-field regime ($E_0\lesssim 1$), the cutoff scales as $\epsilon_{\rm cut,1}=\Delta+\alpha E_0$ with some non-integer constant $\alpha$, see Fig.~\ref{fig:fig2}(d).
The offset $\Delta$, determined from extrapolations $E_0\rightarrow 0$, essentially coincides with the gap size (see arrows in Fig.~\ref{fig:fig2}(d)), which scales linearly with $U$.  
This again leads to the senario that the HHG around the cutoff energy originates from doublon/holon recombination.
In the weak-field regime, the almost unrenormalized spectrum [Fig.~\ref{fig:fig2}(a)] shows that the excited doublons and holons are not localized by the field and thus can move around the lattice to gain kinetic energy (ponderomotive energy $E_{\rm kin}$) and emit this energy in the recombination process. This leads to emission at $n\Omega=\Delta_{\rm gap}+E_{\rm kin}$ in analogy with the three-step model for HHG in atoms and semiconductors \cite{Corkum1993,Lewenstein1994,Vampa2015}. Hence the minimum emission energy from this process is $\Delta_{\rm gap}$.

These results indicate that similar charge dynamics as in semiconductors 
also controls the HHG in MIs, despite the very different nature of these systems.
In semiconductors, when the field is not too strong, the HHG is related to the recombination of itinerant electrons and holes in the valence and conduction bands \cite{Vampa2015}, which yields a linear field dependence of the cutoff energy with an offset.
In the strong-field regime, a quasistatic electric field analysis shows that the HHG originates from transitions among the localized Wannier-Stark states of the conduction and valence electrons, which results in multiple plateaus in the HHG spectrum \cite{Higuchi2014}.

 \begin{figure}[t]
  \centering
    \hspace{-0.cm}
    \vspace{0.0cm}
   \includegraphics[width=86mm]{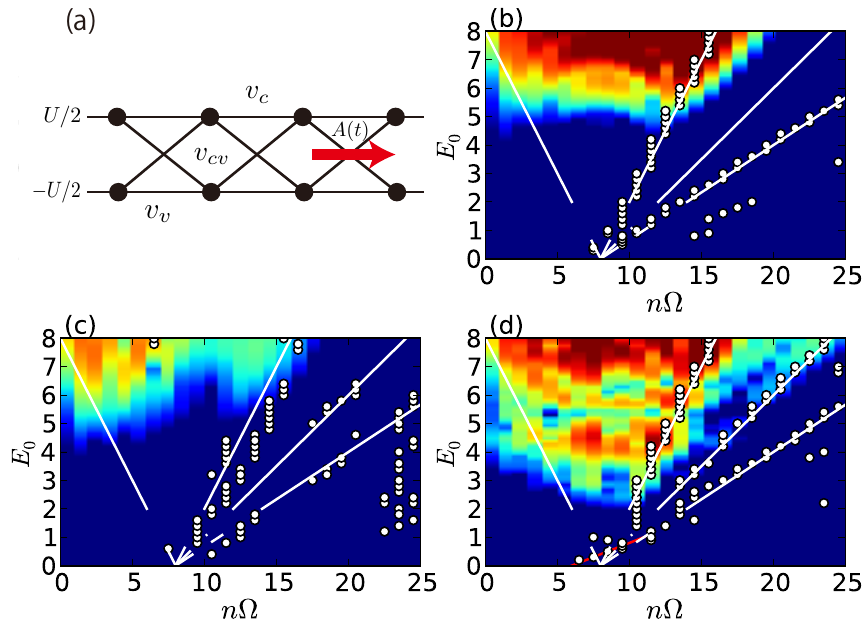} 
  \caption{(a) Schematic picture of the semiconductor model, Eq.~(\ref{eq:H_semicon}). (b)-(d) HHG spectra $I_{hh}$ in the plane of $E_0$ and $n\Omega$. (b) Type 1 semiconductor model. (c) Same model with additional impurity scattering $V_\text{imp}=0.55$. (d) Type 2 semiconductor model. The color scale is the same as in Fig. 2 and $\Omega=0.5,U=8,\beta=2.0,\Gamma=0.06$.}
  \label{fig:fig3}
\end{figure}

In spite of these similarities, we now show that the
 relation between the single particle spectrum and the HHG intensity is very different in MIs and semiconductors.
To this end, 
we study a semiconductor model with a valence band and a conduction band, corresponding to the 
upper and lower Hubbard band, respectively. 
Since in the Hubbard model, the hopping of electrons in MI leads to 
the creation of a doublon/holon pair at neighboring sites, 
we introduce the transfer integral between the different semiconductor orbitals at the neighboring sites.
The resulting Hamiltonian reads 
\small
\begin{align}
H_{\rm semi}(t)
=-&\sum_{\langle i,j\rangle,\alpha}v^\alpha_{ij}(t) c_{i\alpha}^\dagger c_{j\alpha}
-\sum_{\langle i,j\rangle}v^{cv}_{ij}(t) (c^\dagger_{ic}c_{jv}+c^\dagger_{iv}c_{jc})\nonumber\\
&+\sum_{i,\alpha}D_{\alpha}c^\dagger_{i\alpha}c_{i\alpha},\label{eq:H_semicon}
\end{align}
\normalsize
with $D_{\alpha}$ the band center for band $\alpha=\{v, c\}$, see Fig.~\ref{fig:fig3}(a).
 In order to mimic the Hubbard model we choose $D_v=-U/2$ and $D_c=U/2$.
The effect of the electric field is included via the Peierls substitution
and we consider the NESS by attaching a B\"{u}ttiker-type thermal bath.
One reasonable way to determine the hopping parameters is to choose them such that the bands of the semiconductor model show 
a similar dispersion as the Mott insulator. 
In particular, when the Coulomb interaction is large compared to the hopping, 
one may naively expect that the dispersion is given by the Hubbard I (H1) approximation, 
which is based on the atomic-limit self-energy $\Sigma^R(\omega)=\frac{U^2}{4\omega}$ \cite{Hubbard1963}.
Then the dispersion of the upper and lower Hubbard bands becomes $\epsilon_{{\bf k},\pm}=(\epsilon_{\bf k}\pm \sqrt{\epsilon_{\bf k}^2+U^2})/2$, which is reproduced by choosing  $v^c=v^v=v^{cv}=0.5 v$ (we call this ``type 1" model).
The HHG spectrum of the type 1 semiconductor is shown in Fig.~\ref{fig:fig3}(b).
The structure of the HHG spectrum is qualitatively very similar to that from the Hubbard I approximation \cite{supplemental},
and one observes cutoff energies that scale with $U+E_0$ and $U+3E_0$.
However, the model underestimates the HHG spectrum in the weak to intermediate field regime, 
because electron-hole pairs are not efficiently created. 

\begin{figure}[t]
  \centering
    \hspace{-0.cm}
    \vspace{0.0cm}
   \includegraphics[width=86mm]{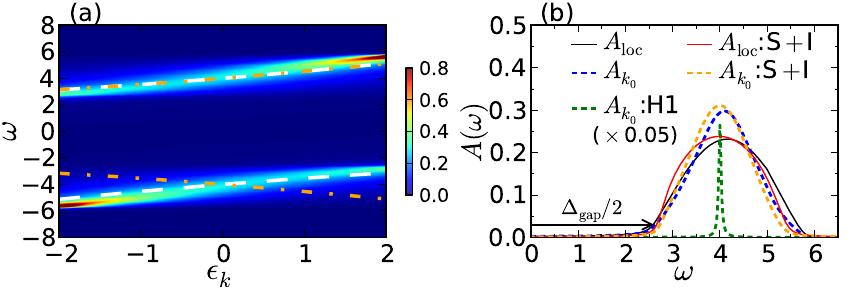} 
  \caption{(a) Momentum dependent spectral function $A(k,\omega)$ of the Mott insulator in equilibrium. 
  The white lines show the peak position predicted by the H1 approximation and the type 1 semiconductor, while the dashed orange lines show the dispersion of the type 2 semiconductor.
  (b) Comparison between the local spectrum $A_{\rm loc}(\omega)$ and $A(k,\omega)$ at $\epsilon_k=0$  $[A_{k_0}]$ obtained within DMFT, H1, and for the type 1 semiconductor with $V_{\rm imp}=0.55$ (S$+$I).
  }
  \label{fig:fig4}
\end{figure}

One major difference between the semiconductor model (or the H1 approximation) and the Hubbard model at finite $U$ 
is that it shows sharp peaks in $A(k,\omega)$, see Fig.~\ref{fig:fig4}.
In the Hubbard model, even though the peak position of  $A(k,\omega)$ at each $\epsilon_{k}$ roughly follows the prediction of the H1 approximation, 
there is a substantial width, comparable to the free electron band width, see Fig.~\ref{fig:fig4}(b).
The incoherence originates from the charge dynamics in a random spin background and does not vanish in the limit $U\rightarrow\infty$ \cite{Metzner1992,supplemental}.
The broadening of the single-particle spectrum can be reproduced in the type 1 semiconductor model by adding impurity effects through the self-energy $\Sigma_{\rm imp}(t,t')=V^2_{\rm imp} G_{\rm loc}(t,t')$, as in~Ref.~\cite{Kemper2013b}, see Fig.~\ref{fig:fig4}(b).
However, as shown in Fig.~\ref{fig:fig3}(c), the resulting HHG spectrum does also not reproduce the HHG spectrum of the MI.
This implies that the strong high-harmonic signal of MIs is not simply related to the 
broadening of the bands. 

Finally, we mention an interesting observation.
The previously proposed three-step models and the strong-field theory have been discussed for semiconductors with a direct gap.
By flipping 
the hopping of the valence band $v^c=-v^v=v^{cv}=0.5 v$ (we call this ``type 2" model),
we realize a direct gap in our semiconductor model. 
In this case, the single particle dispersion is qualitatively different from the dispersion of the MI [Fig.~\ref{fig:fig4}(a)].
However,  as shown in Fig.~\ref{fig:fig3}(d), the resulting HHG spectrum 
reproduces the qualitative features of the HHG spectrum of the MI.
In the weaker field regime the HHG spectrum has a unique plateau and the cutoff energy grows as $\alpha E_0$ with $\alpha>1$, while in the stronger field regime, there emerge multiple plateaus with cutoff  $\epsilon_{{\rm cut},m}=U+mE_0$.
This model also reproduces the characteristic structures i) to iii) observed in the HHG spectrum of MIs. 

{\it Conclusions--}
We revealed the general features of the HHG in wide-gap MIs under continuous AC driving. 
In the strong-field regime, the HHG spectra show multiple plateaus, which is explained by the recombination of localized
doublons and holons at $m$-th nearest neighbor sites. 
In the weak-field regime the HHG spectra show a single plateau, which comes from the recombination of itinerant doublon and holon pairs.
The different nature of MIs and semiconductors is reflected in a very different relation between the high-harmonic spectrum and the single particle spectrum, even though the HHG spectra become qualitatively similar under some conditions.

Strongly correlated systems are known for their high degree of tunability and exotic phases.
In addition, they are a playground for photo-induced phase transitions between different phases.
Therefore, they are an interesting platform to search for new sources of HHG, 
and ultrafast imaging based on HHG might be useful to understand the electron dynamics involved~\cite{Ivanov2018}.
Developing a detailed understanding of the HHG profile in different classes of correlated materials is an interesting topic for future work.

{\it Acknowledgments}
The authors wish to thank D. Gole\v{z}, Z. Lenar\v{c}i\v{c}, M. Sch\"uler, T. Oka and N. Tsuji for fruitful discussions. 
This work was supported by the Swiss National Science Foundation through NCCR MARVEL and the European Research Council through ERC Consolidator Grant 724103. The calculations have been performed on the Beo04 cluster at the University of Fribourg, and the CSCS Dora cluster provided by MARVEL.   

\bibliographystyle{prsty}
\bibliography{Ref,footnote}

\clearpage
\appendix

\section{High-Harmonic Generation in Mott Insulators -- Supplementary Material}

\subsection{Floquet DMFT + NCA}

We briefly discuss the formalism used to solve the DMFT impurity problem. The impurity action can be split into a local term and a hybridization term
\small
\begin{subequations}
\begin{align}
\mathcal{S}_{\rm imp}&=\mathcal{S}_\text{loc}+\mathcal{S}_\text{hyb},\\
\mathcal{S}_\text{loc}&=-i\int_\mathcal{C} dt H_\text{loc}[d^\dagger(t),d(t),t],\\
\mathcal{S}_\text{hyb}&=-i\int_\mathcal{C} dt_1dt_2 \sum_{\sigma_1,\sigma_2} d^\dagger_{\sigma_1}(t_1)\Delta_{\sigma_1,\sigma_2}(t_1,t_2) d_{\sigma_2}(t_2).
\end{align}
\end{subequations}
\normalsize
Here $\mathcal{C}$ indicates the Keldysh contour, and $\int_\mathcal{C}$ is the integral along $\mathcal{C}$. The hybridization function $\Delta(t,t')$ contains a term $\Delta^{\rm DMFT}(t,t')$ related to the hopping of the electron into the surrounding lattice and back, and a term $\Sigma^{\rm bath}(t,t')$ describing the influence of the free electron bath,
\small
\begin{align}
\Delta(t,t')=\Delta^{\rm DMFT}(t,t')+\Sigma^{\rm bath}(t,t').
\end{align}
\normalsize
In a nonequilibrium steady state, $\Delta(t,t')$ satisfies $\Delta(t+\mathcal{T},t'+\mathcal{T})=\Delta(t,t')$.

In this paper, we use the non-crossing approximation (NCA) \cite{Eckstein2010b} to solve the effective impurity model with a time periodic hybridization function and an electron bath.
NCA is the lowest-order self-consistent strong coupling expansion in the hybridization function.
It can be formulated by introducing pseudo-particles for each local state of the impurity site.
In the case of the single-band Hubbard model, the local states are $|\!\uparrow\downarrow\rangle,|\!\uparrow\rangle, |\!\downarrow\rangle,|{\rm vac}\rangle$ and we introduce creation operators 
$\hat{a}^\dagger_{\uparrow\downarrow},\hat{a}^\dagger_{\uparrow},\hat{a}^\dagger_{\downarrow},\hat{a}^\dagger_{0}$ for each state.
The operator $\hat{a}^\dagger_{\uparrow\downarrow}$ can be interpreted as the creation operator of a doublon, while $\hat{a}^\dagger_{0}$ as that of a holon.
Here $\hat{a}^\dagger_m$ is fermionic (bosonic) when the local state ($|m\rangle$) represents an odd (even) number of fermions. 
The total number of pseudo-particles is $Q\equiv \sum_m a^\dagger_m a_m$.
The physical Hilbert space in the Fock space of the pseudo-particles is limited to the subspace $Q=1$.
Using the pseudo-particle operators, we introduce 
\small
\begin{subequations}
\begin{align}
\tilde{d}^\dagger_\uparrow=a^\dagger_\uparrow a_0-a^\dagger_{\uparrow\downarrow} a_\downarrow,\\
\tilde{d}^\dagger_\downarrow=a^\dagger_\downarrow a_0+a^\dagger_{\uparrow\downarrow} a_\uparrow,
\end{align}
\end{subequations}
\normalsize
which are identical to the original electron operators in the physical space.
Using these identities, we can express the impurity action in terms of pseudo-particle operators: $\mathcal{S}_\text{imp}[\tilde{d},\tilde{d}^\dagger]\equiv \tilde {\mathcal{S}}_\text{imp}[a,a^\dagger]$.

Our goal is to evaluate the physical Green's function of the impurity site.
In terms of pseudo-particles, it can be regarded as a two particle Green's function.
One can express it as a combination of single-particle Green's functions of the pseudo-particles.
Reflecting the fact that the physically relevant space satisfies $Q=1$, we introduce the (projected) pseudo-particle Green's function as
\small
\begin{align}
\mathcal{G}_{mm'}&(t,t')=\nonumber\\
&\theta_\mathcal{C}(t,t')(-i){\rm Tr}_{Q=0}[T_\mathcal{C} a_m(t)a^\dagger_{m'}(t') \exp(\tilde{\mathcal S}_\text{imp})]\nonumber\\
+&\theta_\mathcal{C}(t',t)(-i){\rm Tr}_{Q=1}[T_\mathcal{C} a_m(t)a^\dagger_{m'}(t') \exp(\tilde{\mathcal S}_\text{imp})].
\end{align}
\normalsize
Here $T_\mathcal{C}$ is the contour ordering operator and $\theta_\mathcal{C}(t,t')$ is the Heaviside function on the contour.
The Dyson equation for the pseudo-particle Green's function is
\small
\begin{align}
[i\partial_t-h(t)]\mathcal{G}(t,t')-\int_{\mathcal C,t'<\bar{t}<t} d\bar{t} \Sigma(t,\bar{t})\mathcal{G}(\bar{t},t')=\delta_\mathcal{C}(t,t').\label{eq:Dyson_eq_pp}
\end{align}
\normalsize
Here $\delta_\mathcal{C}(t,t')$ is the delta function on the Keldysh contour $\mathcal{C}$, and ``$\mathcal C,t'<\bar{t}<t$" indicates that the time arguments $t',\bar{t}$ and $t$ are in cyclic order along $\mathcal{C}$~\cite{Aoki2013}. Here we use the matrix form of the Green's function in terms of local states, $h(t)$ indicates the local Hamiltonian and $\Sigma$ is the pseudo-particle self-energy\cite{Eckstein2010b}.

We define the components of the pseudo-particle Green's function as \cite{Aoki2013}
\small
\begin{subequations}
\begin{align}
\mathcal{G}^>_{mm'}(t,t')&=-i{\rm Tr}_{Q=0}[T_\mathcal{C} a_m(t)a^\dagger_{m'}(t') \exp(\tilde{\mathcal{S}}_\text{imp})],\\
\mathcal{G}^<_{mm'}(t,t')&=-i{\rm Tr}_{Q=1}[T_\mathcal{C} a_m(t)a^\dagger_{m'}(t') \exp(\tilde{\mathcal{S}}_\text{imp})],\\
\mathcal{G}^R(t,t')&=\theta(t-t') \mathcal{G}^>(t,t'),\\
\mathcal{G}^A(t,t')&=-\theta(t'-t) \mathcal{G}^>(t,t'),\\
\mathcal{G}^K(t,t')&=\mathcal{G}^>(t,t'),
\end{align}
\end{subequations}
\normalsize
and the same definition is applied to the pseudo-particle self-energies.
Here $>,<,R,A,K$ represent the greater, lesser, retarded, advanced and Keldysh parts respectively,
We note that different from the normal Green's function, the retarded, advanced and Keldysh parts are not independent.
 \begin{figure*}
  \centering
    \hspace{-0.cm}
    \vspace{0.0cm}
   \includegraphics[width=150mm]{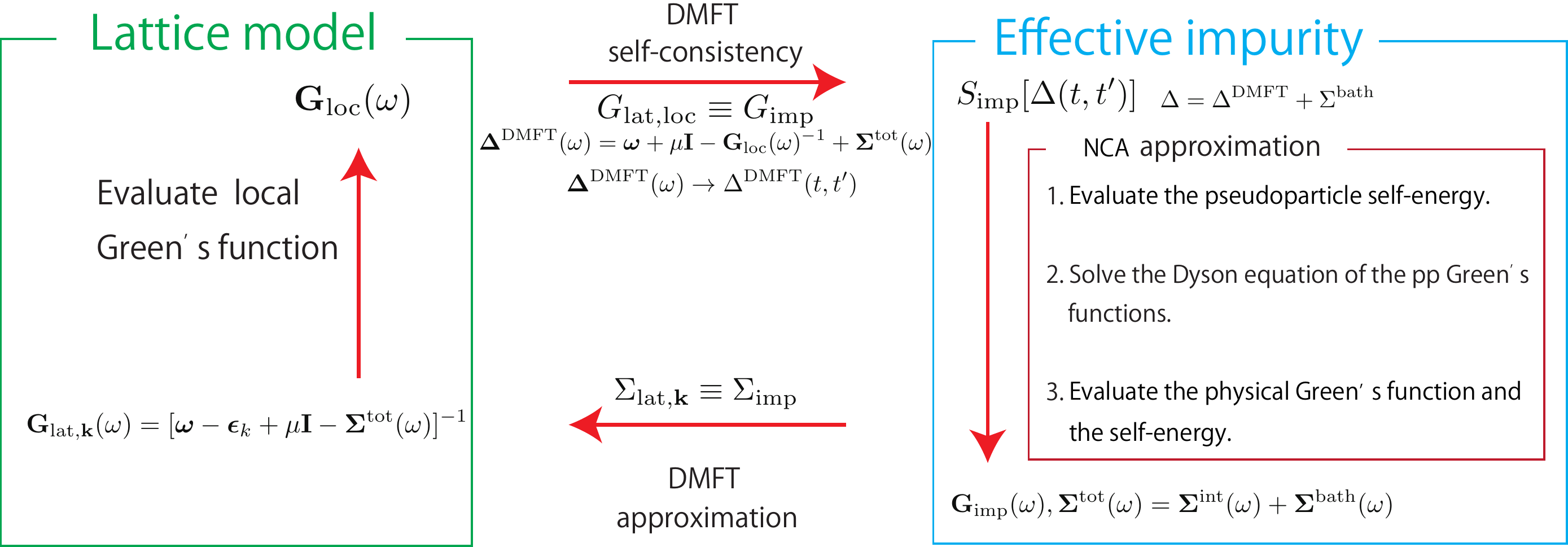} 
  \caption{Self-consistency loop for the Floquet DMFT implemented with the strong coupling expansion (NCA) as an impurity solver. 
  Boldified letters indicates (${\bf F}(\omega)$) the Floquet representation of a certain function ($F(t,t')$). See Ref.~\cite{Aoki2013} for the detailed definition.
  }
  \label{fig:flow}
\end{figure*}
For the normal Green's function, one can obtain a transparent expression of the Dyson equation by representing the Green's function 
in the Larkin-Ovchinnikov form, which consists of $G^R,G^K$ and $G^A$~\cite{Aoki2013,Murakami2017}.
For the pseudo-particle Green's functions, it turns out that the Dyson equation can be expressed in a similar manner if 
we consider the matrix consisting of $\mathcal{G}^R,\mathcal{G}^<$ and $\mathcal{G}^A$,
\small
\begin{align}
\underline{\mathcal{G}}(t,t')&\equiv
\begin{bmatrix}
\mathcal{G}^{R}(t,t')&\mathcal{G}^{<}(t,t')\\
0&\mathcal{G}^{A}(t,t')
\end{bmatrix},
\end{align}
\normalsize
\small
\begin{align}
[i\partial_t-h(t)]\underline{\mathcal{G}}(t,t')-\int^\infty_{-\infty} d\bar{t}\underline{\Sigma}(t,\bar{t})\underline{\mathcal{G}}(\bar{t},t')=\delta(t-t')\underline{I}.
\label{eq:Dyson_Keldysh}
\end{align}
\normalsize
Therefore, one can solve the Dyson equation for the pseudo particles using the Floquet representation 
in the same manner as for the usual Green's function~\cite{Aoki2013,Murakami2017},
\small
\begin{align}
[\underline{\mathcal{G}}_0(\omega)^{-1}-\underline{\boldsymbol \Sigma}(\omega)]\underline{\mathcal {G}}(\omega)&=\underline{\bf I}.
\end{align}
\normalsize

So far we have explained the general framework of the hybridization expansion impurity solver formulated in terms of pseudo-particles.
In the following, we explicitly show the expressions for the NCA.
The pseudo-particle self-energy is evaluated by an expansion in the hybridization.
NCA gives the lowest order (self-consistent) contribution.
In the Hubbard model considered here, only the diagonal components are finite,
because $\Delta$ is diagonal, 
and they can be expressed as
\small
\begin{subequations}
\begin{align}
\Sigma^{NCA}_{0}(t,t')&=-i\sum_\sigma \Delta_{\sigma}(t',t)\mathcal{G}_\sigma(t,t'),\\
\Sigma^{NCA}_{\sigma}(t,t')&=i\Delta_{\sigma}(t,t')\mathcal{G}_0(t,t')-i\Delta_{\bar{\sigma}}(t',t)\mathcal{G}_{\udarrow}(t,t'),\\
\Sigma^{NCA}_{\udarrow}(t,t')&=i\sum_\sigma \Delta_{\sigma}(t,t')\mathcal{G}_{\bar{\sigma}}(t,t').
\end{align}
\end{subequations}
\normalsize
The expression of the physical impurity Green's function ($G_{{\rm imp},\sigma}(t,t')=-i\langle T_{\mathcal{C}} d_{\sigma}(t)d^\dagger_{\sigma}(t')\rangle$) is
\small
\begin{align}\label{eq:G_nca}
G_{{\rm imp},\sigma}(t,t')=i[\mathcal{G}_\sigma(t,t')\mathcal{G}_0(t',t)-\mathcal{G}_{\udarrow}(t,t')\mathcal{G}_{\bar{\sigma}}(t',t)]/\tilde{Q},
\end{align} 
\normalsize
where $\tilde Q\equiv i\sum_m (-1)^m \mathcal{G}^<_{mm}(t,t)$.

In DMFT, we identify the impurity Green's function with the local Green's function of the lattice model, $G_{\rm loc,\sigma}(t,t')=-i\langle T_{\mathcal{C}} c_{i,\sigma}(t)c^\dagger_{i,\sigma}(t')\rangle$.
The self-consistency loop of the Floquet DMFT implemented with the NCA impurity solver is illustrated in  Fig.~\ref{fig:flow}.
We note that the NCA part can be replaced by higher-order schemes of the strong-coupling expansion such as the one-crossing approximation \cite{Eckstein2010b}.

\subsection{Current contributions in the Mott insulator}
In order to identify the origin of the current and relate it to the doublon/holon dynamics, we introduce the pseudo-particles for each site and an operator 
\small
\begin{align}\label{eq:pp_lattice}
\tilde{c}^\dagger_{i\sigma}&=a^\dagger_{i\sigma}a_{i0}+(-)^\sigma a^\dagger_{i\uparrow\downarrow}a_{i\bar{\sigma}}
&\equiv D^{(1)\dagger}_{i\sigma}+(-)^\sigma D^{(2)\dagger}_{i\sigma},
\end{align}
\normalsize
which is identical to the original electron creation operator $c^\dagger_{i\sigma}$ in the physical space.
Here $(-)^\uparrow=-1,\;(-)^\downarrow=1$.
The physical space is gained by the projection to $a^\dagger_{i\uparrow\downarrow}a_{i\uparrow\downarrow}+a^\dagger_{i\uparrow}a_{i\uparrow}+a^\dagger_{i\downarrow}a_{i\downarrow}+a^\dagger_{i0}a_{i0}=1$ for each $i$.
With these operators, the polarization along the direction of the external field is 
\small
\begin{align}
\hat{P}=q\sum_i {\bf e}_0\cdot {\bf r}_i (a^\dagger_{i\uparrow\downarrow}a_{i\uparrow\downarrow}-a^\dagger_{i0}a_{i0}+1) 
\end{align}
\normalsize
and 
\small
\begin{align}
&H(t)=H_{\rm loc}\nonumber\\
&-\sum_{i,j,\sigma}v_{ij}(t) \left( a^\dagger_{i\sigma}a_{i0}+(-)^\sigma a^\dagger_{i\uparrow\downarrow}a_{i\bar{\sigma}} \right) 
\left( a^\dagger_{j0}a_{j\sigma}+(-)^\sigma a^\dagger_{j\bar{\sigma}} a_{j\uparrow\downarrow} \right).
\end{align}
\normalsize
$H_{\rm loc}$ at half-filling is $H_{\rm loc}=\frac{U}{2}\sum_i(a^\dagger_{i\uparrow\downarrow}a_{i\uparrow\downarrow}+a^\dagger_{i0}a_{i0})$.
The kinetic term can be decomposed into three parts,
\small
\begin{subequations}
\begin{align}
H_{\rm kin,h}(t)&\equiv-\sum_{i,j,\sigma} v_{ij}(t) a^\dagger_{i\sigma}a_{i0}  a^\dagger_{j0}a_{j\sigma},\label{eq:kin_h}\\
H_{\rm kin,d}(t)&\equiv-\sum_{i,j,\sigma} v_{ij}(t) a^\dagger_{i\uparrow\downarrow}a_{i\sigma} a^\dagger_{j\sigma} a_{j\uparrow\downarrow},\label{eq:kin_d}\\
H_{\rm kin,dh}(t)&\equiv-\sum_{i,j,\sigma} v_{ij}(t) (-)^\sigma (a^\dagger_{i\uparrow\downarrow}a_{i\bar{\sigma}}a^\dagger_{j0}a_{j\sigma}+a^\dagger_{i\sigma}a_{i0}a^\dagger_{j\bar{\sigma}} a_{j\uparrow\downarrow}).\label{eq:kin_dh} 
\end{align}
\end{subequations}
\normalsize
The first term represents holon hopping from $i$ to $j$, the second term doublon hopping from $j$ to $i$ and the third term doublon-holon pair creation and recombination between $i$ and $j$.

Since the current is $j(t)=\partial_t P(t)$, we can identify the contributions from these three processes.
The contribution from the holon hopping can be expressed as 
\small
\begin{align}
\hat{j}_{\rm hop,h}&=-i[\hat{P},H_{\rm kin,h}(t)]\nonumber\\
&=iq\sum_{i,j,\sigma} v_{ij}(t) ({\bf e}_0\cdot {\bf r}_{i-j}) a^\dagger_{i\sigma}a_{i0}  a^\dagger_{j0}a_{j\sigma},
\end{align}
\normalsize
while the contribution from the doublon hopping is 
\small
\begin{align}
\hat{j}_{\rm hop,d}&=-i[\hat{P},H_{\rm kin,d}(t)]\nonumber\\
&=iq\sum_{i,j,\sigma} v_{ij}(t) ({\bf e}_0\cdot {\bf r}_{i-j}) a^\dagger_{i\uparrow\downarrow}a_{i\sigma} a^\dagger_{j\sigma} a_{j\uparrow\downarrow}.
\end{align}
\normalsize
Hence the total current coming from hopping of doublons and holons is $\hat{j}_{\rm hop}\equiv \hat{j}_{\rm hop,d}+\hat{j}_{\rm hop,h}$.
The contribution from the recombination/creation of a doublon-holon pair is 
\small
\begin{align}
\hat{j}_{\rm rc}=&-i[\hat{P},H_{\rm kin,dh}(t)]=iq\sum_{i,j,\sigma} v_{ij}(t) ({\bf e}_0\cdot {\bf r}_{i-j}) \nonumber\\
&\times (-)^\sigma (a^\dagger_{i\uparrow\downarrow}a_{i\bar{\sigma}}a^\dagger_{j0}a_{j\sigma}+a^\dagger_{i\sigma}a_{i0}a^\dagger_{j\bar{\sigma}} a_{j\uparrow\downarrow}). 
\end{align}
\normalsize
Conceptually, the first two terms are analogous to the contribution from the intraband 
motion 
of the electrons and holes in semiconductors, while the last term corresponds to the creation and recombination of electrons and holes in  semiconductors (polarization current). These considerations are applicable in any dimension $d\ge 1$. 

\subsection{Generalized tunneling formula}

Here we introduce the generalized tunneling formula for the current in nonequilibrium steady states (NESSs), 
which helps us to understand the physical processes involved.
In Ref.~\cite{Lee2014}, the authors have derived a tunneling formula for NESSs driven by a DC field,
which is justified when the hopping is small enough compared to the interaction.
In practice, it works quantitatively well as shown in Fig.~5(b) of Ref.~\cite{Lee2014}.
Here we briefly explain how to generalize the idea to AC fields.
A more detailed discussion and analysis of the formula is presented in Ref.~\cite{Murakami2018}.

First, we select one direction in the hyper-cubic lattice ($x$) and 
regard the system as a stack of $(d-1)$-dimensional slabs, which are alined in the $x$ direction.
The Hamiltonian can now be expressed as $\hat{H}(t)=\hat{H}_{\perp}(t)+\hat{V}_{x}(t)$, where $\hat{V}_{x}(t)$
describes the transfer integrals along the $x$ direction (junctions between slabs), and $\hat{H}_{\perp}(t)$ the $(d-1)$-dimensional slabs.

In the Floquet steady state, initial correlations are washed out because of the heat bath.
Therefore, one can prepare the steady state of the full system by starting from a steady state of $\hat{H}_{\perp}$, where all slabs are disconnected, and adiabatically switching on $\hat{V}_{x}$.
When $U$ is large, the effect of $\hat{V}_{x}$ can be treated perturbatively.
Here we consider the linear contribution to the wave function and evaluate the current in the $x$ direction.
The first order correction of $\hat{V}_x$ to the state is 
\small
\begin{align}
|\Psi(t)\rangle\simeq|\Psi^{(0)}(t)\rangle-i\int^t_{-\infty}d\bar{t}\hat{\mathcal{U}}_0(t,\bar{t}) \hat{V}_x |\Psi^{(0)}(\bar{t})\rangle.
\end{align} 
\normalsize
Here $\hat{\mathcal{U}}_0(t,t')=\mathcal{T}\exp[-i\int^t_{t'} d\bar{t} \hat{H}_{\rm tot,0}(\bar{t})]$ for $t>t'$
and $\mathcal{T}$ is the 
time-ordering operator.
Hence, the current along the $x$ direction is 
\small
\begin{align}
j_x(t)=-i\int^t_{t_0}d\bar{t}s\langle \Psi^{(0)}(t) | \hat{j}_x\hat{\mathcal{U}}_0(t,\bar{t}) \hat{V}_x |\Psi^{(0)}(\bar{t})\rangle +H.c.
\end{align}
\normalsize

We can connect this expression to the local Green's functions by using the following conditions: 
i) the Floquet steady state should be a mixed state of all Floquet states, 
ii) the density of states (DOS) of each slab can be approximated with the full $d$-dimensional bulk result since $d$ is large,
iii) we only consider the contribution to the current at a certain junction by an electron that went  through the same junction. 
We then obtain the generalized tunneling formula that connects the local Green's function and the current in the NESS,
\small
\begin{align}
j_{\rm tun}(t)=-qv^{*2}&{\rm Re}\Bigl[\int_{-\infty}^t d\bar{t} \{G_{\rm loc}^<(\bar{t},t)G_{\rm loc}^>(t,\bar{t})\nonumber\\
&-G_{\rm loc}^>(\bar{t},t)G_{\rm loc}^<(t,\bar{t})\}e^{-i\int^t_{\bar{t}}dt' E(t')}\Bigl]. \label{eq:j_tun_tot}
\end{align}
\normalsize
Here we use $v=\frac{v^*}{2\sqrt{d}}$ and consider the contribution from all directions.
This formula is applicable both to the DC field case (where it reduces to Eq. (96) in Ref.~\cite{Lee2014}) and the AC field case.
It turns out that this formula is qualitatively and even quantitatively 
valid in the parameter regime used in this paper \cite{Murakami2018}.

 \begin{figure}[t]
  \centering
    \hspace{-0.cm}
    \vspace{0.0cm}
   \includegraphics[width=84mm]{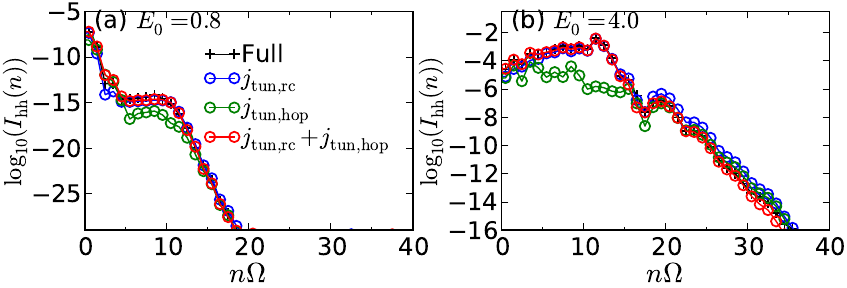} 
  \caption{Comparison of the HHG intensity evaluated from the current involving different processes in the weak-field regime (a) and the strong-field regime (b).
  ``Full" indicates the full FDMFT+NCA calculation, $j_{\rm rc}$ is the current from the recombination process, and $j_{\rm hop}$ is the current from the hopping process of doublons and holons. 
  $j_{\rm rc}$ and $j_{\rm hop}$ are evaluated from the generalized tunneling formula. Here $U=8.0,\beta=2.0,\Gamma=0.06$ and $W_{\rm bath}=5$.}
  \label{fig:comp_hhg}
\end{figure}

Next, we explain how to evaluate the contributions from the two different processes $\hat{j}_{\rm hop}$ and $\hat{j}_{\rm rc}$
defined in the previous section.
To this end we introduce the Green's function for the operators $D$ defined in Eq.~(\ref{eq:pp_lattice}) as 
\small
\begin{align}\label{eq:Green_D}
G_{{\rm loc},ll',\sigma}(t,t')=-i\langle \mathcal{T}_{\mathcal{C}}D_{i\sigma}^{(l)}(t) D_{i\sigma}^{(l')\dagger}(t')\rangle.
\end{align}
\normalsize
If the DMFT equations are solved with a strong-coupling (NCA, OCA, etc.) impurity solver, one can directly evaluate this quantity. 
In particular, NCA implies that $G_{{\rm loc},ll',\sigma}(t,t')=0$ when $l\neq l'$, and 
 $G_{{\rm loc},11,\sigma}(t,t')$ and $G_{{\rm loc},22,\sigma}(t,t')$ correspond to the first and second terms in Eq.~(\ref{eq:G_nca}), respectively.

Applying the above argument for the total current to $\hat{j}_{\rm hop}$ and $\hat{j}_{\rm rc}$, we obtain for FDMFT+NCA
\small
\begin{align}\label{eq:j_tun_rec}
j_{\rm tun, cr}(t)=&-\frac{qv^{*2}}{2} {\rm Re}  \int^t_{-\infty}
\sum_{\sigma,l}\bigl[G^<_{{\rm loc},ll,\sigma}(\bar{t},t)
G^>_{{\rm loc},\bar{l}\bar{l},\sigma}(t,\bar{t})\nonumber\\
&\;\;\;-G^>_{{\rm loc},ll,\sigma}(\bar{t},t)G^<_{{\rm loc},\bar{l}\bar{l},\sigma}(t,\bar{t})
\bigl]e^{-i\int^t_{\bar{t}}dt' E(t')},
\end{align}
\normalsize
and 
\small
\begin{align}\label{eq:j_tun_hop}
j_{\rm tun, hop}(t)
=&-\frac{qv^{*2}}{2} {\rm Re} \int^t_{-\infty}
\sum_{\sigma,l}\bigl[G^<_{{\rm loc},ll,\sigma}(\bar{t},t)G^>_{{\rm loc},ll,\sigma}(t,\bar{t})\nonumber\\
&\;\;\;-G^>_{{\rm loc},ll,\sigma}(\bar{t},t)G^<_{{\rm loc},ll,\sigma}(t,\bar{t})
\bigl]e^{-i\int^t_{\bar{t}}dt' E(t')}.
\end{align}
\normalsize
We note that $j_{\rm tun}(t)=j_{\rm tun, rc}(t)+j_{\rm tun, hop}(t)$.

In Fig.~\ref{fig:comp_hhg}, we compare the contributions to the HHG from the recombination ($j_{\rm rc}$) and hopping ($j_{\rm hop}$),
which are evaluated with Eq.~(\ref{eq:j_tun_rec}) and Eq.~(\ref{eq:j_tun_hop}).
The sum of both contributions is almost identical to the exact result,  
which demonstrates the validity of the formula.
One can see that in both cases, the dominant contribution is coming from the recombination/creation process both in the weak-field and strong-field regimes,
which quantitatively supports our statements in the original manuscript. 
We also note that the contribution from the hopping process roughly follows that of the recombination process.
This is not a strange result because these processes cannot be fully decoupled 
since the wave function is not fully localized at a given site.
A similar effect has been reported in a semiconductor study.
There, the 
effect of the recollision of electrons and holes 
also partially appears in the intraband current, see Eq.~(4) and Fig.~1 of Ref.~\cite{Vampa2014}.

\subsection{Spectral function}

\begin{figure}[t]
  \centering
    \hspace{-0.cm}
    \vspace{0.0cm}
   \includegraphics[width=68mm]{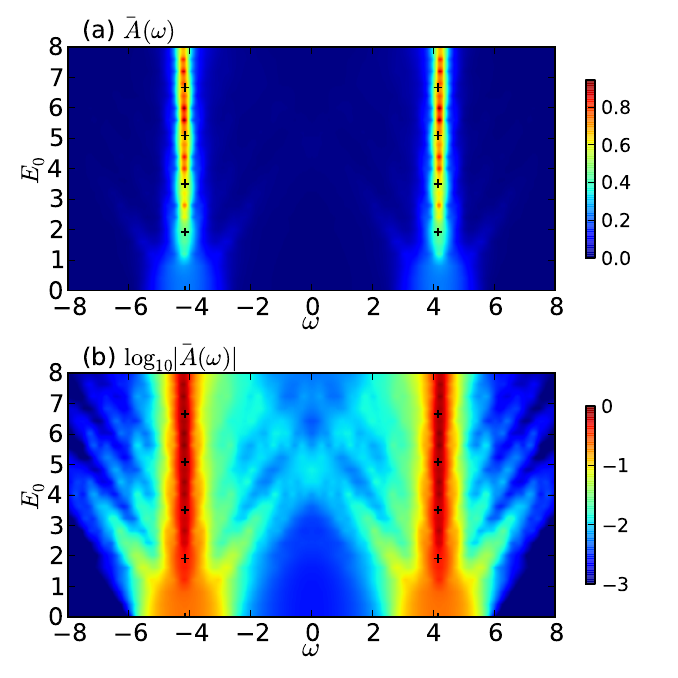} 
  \caption{Time-averaged local spectral function for $U=8.0,\beta=2.0,\Gamma=0.06,W_{\rm bath}=5.0$ in the space of $\omega$ and field strength $E_0$ on a linear scale (a) and a logarithmic scale (b). The maxima of the hopping renormalization factor $|J_0(E_0/\Omega)|$ are indicated by `$+$' marks.}
  \label{fig:supp_fig5}
\end{figure}

The field dependence of the mobility of the charge carriers (doublons and holons) manifests itself in the time-averaged local spectral function $\bar A(\omega)\equiv -\frac{1}{\pi}{\rm Im} \bar{G}^R_{\rm loc}(\omega)$.
Here $\bar{G}^R_{\rm loc}(\omega)=\frac{1}{\mathcal{T}}\int^{\mathcal{T}}_0 dt_{\rm av}\int dt_{\rm r}G^R_{\rm loc}(t_{\rm r};t_{\rm av})e^{i\omega t_{\rm r}}$. Here $t_{\rm r}$ is the relative time and $t_{\rm av}$ is the average time.
We illustrate the dependence of $\bar A$ on the field strength in Fig.~\ref{fig:supp_fig5} for the parameters used in the main text.
In the weak-field regime ($E_0\lesssim1$), 
the width of the Hubbard bands is not much renormalized and remains about 2 (width at half-maximum).
In the strong-field regime ($E_0\gtrsim2$), the width of the Hubbard bands is substantially decreased and there emerge clear side bands besides the main Hubbard bands, whose peak positions linearly scale with the field strength as $\pm \frac{U}{2} \pm \gamma E_0$.
These are manifestation of the localization of charge carriers. The side bands corresponds to the Wannier-Stark states
that have been observed in the Hubbard and Holstein-Hubbard model under DC fields \cite{Lee2014,Werner2015}.
Reflecting the oscillating nature of the AC field with maximum field strength $E_0$, the coefficient $\gamma$ which determines the slope of the Wannier-Stark sidebands is smaller than 1.
We also note that the additional sidebands branch off at the field strengths corresponding to the maxima in the hopping renormalization factor $|J_0(E_0/\Omega)|$, which are indicated by `$+$' marks.

Now next we discuss the width of the peak in the momentum-resolved spectral function $A(k,\omega)$ in equilibrium.
Our analytic and numerical studies show that NCA yields a finite width of the momentum-resolved spectral function $A(k,\omega)$, of the order of the hopping $v$, even when the interaction is much larger than the hopping. 
We note that these results are consistent with an analytical study of the $t$-$J$ model~\cite{Metzner1992}, which is obtained in the large $U$ limit from the Hubbard model. 

For the analytic study, we consider a one-shot (bare) NCA.
This calculation involves the following steps:
\begin{enumerate}
\item Approximate the hybridization function by the local Green's function in the atomic limit: $\Delta\simeq v^2 G_{\rm loc,atom}$. 
\item Evaluate the pseudo-particle self-energy by $\Delta\cdot \mathcal{G}_{m}^{(0)}$, where $\mathcal{G}_{m}^{(0)}$ is the bare pseudo-particle Green's function in the atomic limit.
\item Evaluate the self-energy of the physical Green's function and use it in the lattice Dyson equation to calculate $G_k$.
\end{enumerate}

In equilibrium, the hybridization function in the atomic limit approximation is 
\small
\begin{align}
\Delta^<(t)&=-\Delta^>(-t)=\frac{iv^2}{2}\left[f\bigl(\tfrac{U}{2}\bigl)e^{-it\frac{U}{2}}+f\bigl(-\tfrac{U}{2}\bigl)e^{it\frac{U}{2}}\right],
\end{align}
\normalsize
%
while 
the pseudo-particle Green's function in the atomic limit is 
\small
\begin{subequations}
\begin{align}
\mathcal{G}^{>(0)}_{m}(t)&=-ie^{-i t \epsilon_m}.
\end{align}
\end{subequations}
\normalsize
Here $f(\omega)$ is the Fermi distribution function at inverse temperature $\beta$ and   $\epsilon_{\uparrow \downarrow}=\epsilon_0=0$, $\epsilon_{\sigma}=-U/2$ at half-filling. 

The pseudo-particle self-energy for the retarded part evaluates to 
\small
\begin{subequations}
\begin{align}
\Sigma^{R}_{0}(t)&=\Sigma^{R}_{\udarrow}(t)=-i\sum_\sigma \Delta_{\sigma}^<(-t)\mathcal{G}_{\sigma}^{R(0)}(t)\nonumber\\
&=-iv^2\theta(t)(f\bigl(\tfrac{U}{2}\bigl)e^{iUt}+f\bigl(-\tfrac{U}{2}\bigl)),\\
\Sigma^{R}_{\sigma}(t)&=i\Delta_{\sigma}^>(t)\mathcal{G}_{0}^{R(0)}(t)-i\Delta_{\bar{\sigma}}^<(-t)\mathcal{G}^{R(0)}_{\udarrow}(t)\nonumber\\
&=-i v^2 \theta(t) \left[f\bigl(-\tfrac{U}{2}\bigl)e^{-i\frac{U}{2}t}+f\bigl(\tfrac{U}{2}\bigl)e^{i\frac{U}{2}t}\right].
\end{align}
\end{subequations}
\normalsize

By solving the Dyson equation for the retarded part in Fourier space, we obtain
\small
\begin{subequations}\label{eq:pp_Green}
\begin{align}
\mathcal{G}_{0}^R(\omega)&=\frac{\omega(\omega+U)}{\omega^2(\omega+U)-v^2(\omega+U)+v^2f(\frac{U}{2})U}\nonumber\\
&\equiv \sum_{\alpha=-1,0,1}\frac{A_{0,\alpha}}{\omega-\epsilon_{0,\alpha}},\\
\mathcal{G}_{\sigma}^R(\omega)&=\frac{\omega^2-U^2/4}{(\omega+\frac{U}{2})^2(\omega-\frac{U}{2})-v^2(\omega+\frac{U}{2})+v^2f(\frac{U}{2})U}\nonumber\\
&\equiv \sum_{\alpha=-1,0,1}\frac{A_{\sigma,\alpha}}{\omega-\epsilon_{\sigma,\alpha}}.
\end{align}
\end{subequations}
\normalsize
We note that $v^2f(\frac{U}{2})U$ becomes exponentially small when the temperature is small or $U$ is large.
One can see that $\sum_{\alpha}A_{m,\alpha}=1$, $A_{m,\alpha}\in {\bf R}$, $\epsilon_{0,\pm1}\simeq \pm v$, $\epsilon_{0,0}\simeq -U$, $\epsilon_{\sigma,0}\simeq U$ and $\epsilon_{\sigma,\pm1}\simeq \pm \tilde{U}$. Here $\tilde{U}=\sqrt{U^2+v^2/4}$.

From this it follows that 
\small
\begin{align}
\mathcal{G}_{m}^>(t)&=-i\sum_{\alpha}A_{m,\alpha} e^{-i\epsilon_{m,\alpha}t},
\end{align}
and by the substitution $it\rightarrow  \tau$ in the greater component, we obtain
\begin{align}
\mathcal{G}_m^M(\tau)=-\sum_{\alpha} A_{m,\alpha} e^{-\tau \epsilon_{m,\alpha}} \text{  (for $\tau>0$)}.
\end{align}
\normalsize
Now we evaluate the local Green's function on the Matsubara axis.
The physical Green's function is 
\small
\begin{align}
&G^M(\tau)=-\left[ \mathcal{G}_\sigma^M(\tau)\mathcal{G}_0^M(-\tau) - \mathcal{G}_{\udarrow}^M(\tau) \mathcal{G}_{\bar{\sigma}}^M(-\tau) \right]/\tilde{Q}\nonumber\\
=&-\frac{1}{\tilde{Q}} \sum_{\alpha,\gamma} A_{\sigma,\alpha} A_{0,\gamma} \Big[e^{-\beta \epsilon_{0,\gamma}}e^{-\tau(\epsilon_{\sigma,\alpha}-\epsilon_{0,\gamma})}\nonumber\\
&+e^{-\beta \epsilon_{\sigma,\alpha}}e^{\tau(\epsilon_{\sigma,\alpha}-\epsilon_{0,\gamma})}\Big]
\end{align}
\normalsize
and
\small
\begin{align}
\tilde{Q}&=-\sum_m (-)^m \mathcal{G}_0^M(-0^+)\nonumber\\
&=2\sum_\alpha A_{0,\alpha} e^{-\epsilon_{0,\alpha}\beta}+2\sum_{\alpha} A_{\sigma,\alpha} e^{-\epsilon_{\sigma,\alpha}\beta}.
\end{align}
\normalsize
Here, the negative $\tau$ component of the pseudo-particle Green's function can be obtained by the usual \mbox{(anti-)periodic} relation 
$\mathcal{G}_m^M(-\tau)=(-)^m\mathcal{G}_m^M(\beta-\tau)$.

After expressing $G^M$ in Matsubara frequency space and performing the analytic continuation $i\omega_n\rightarrow \omega$, 
we obtain 
\small
\begin{align}\label{eq:one_nca_green}
G^R(\omega)&=\frac{1}{\tilde{Q}} \sum_{\alpha,\beta} 
 \frac{A_{\sigma,\alpha} A_{0,\beta} (e^{-\beta \epsilon_{0,\beta}}+e^{-\beta \epsilon_{\sigma,\alpha}})}{\omega^2-(\epsilon_{\sigma,\alpha}-\epsilon_{0,\beta})^2}.
\end{align}
\normalsize
In the limit of $\beta\rightarrow\infty$ this simplifies to 
\small
\begin{align}
G^R(\omega)=\frac{1}{4}\sum_{a=\pm 1,b=\pm 1}\frac{1}{\omega+a \tilde{U}+b v}.
\end{align}
\normalsize

 \begin{figure}[t]
  \centering
    \hspace{-0.cm}
    \vspace{0.0cm}
   \includegraphics[width=86mm]{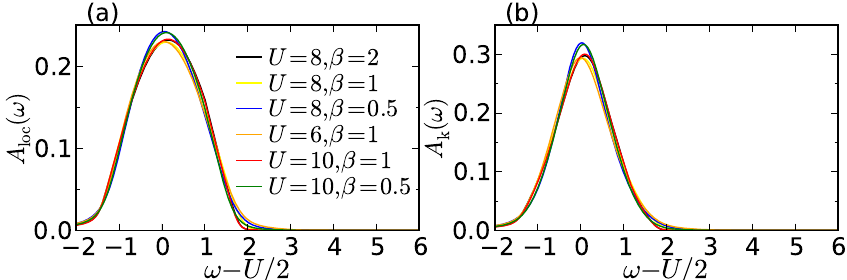} 
  \caption{(a) Local spectral function $A_{\rm loc}(\omega)$ and  
  (b) momentum-dependent spectral function $A(k,\omega)$ at $\epsilon_k=0$ in equilibrium for various $U$ and temperatures. Here $\Gamma=0.06$.}
  \label{fig:supp_fig6}
\end{figure}

From $\Sigma^R(\omega)=\omega-\Delta^R(\omega)-G^R(\omega)^{-1}$, the lattice Green's function becomes
\small
\begin{align}
G^R(k,\omega)^{-1}=&\omega-\epsilon_k-\Sigma^R(\omega)\nonumber\\
=&-\epsilon_k+v^2\frac{\omega}{\omega^2-\frac{U^2}{4}}\nonumber\\
&+\frac{2}{\omega}\frac{(\omega^2-(\frac{\tilde{U}}{2}+v)^2)(\omega^2-(\frac{\tilde{U}}{2}-v)^2)}{2\omega^2-\frac{\tilde{U^2}}{2}-2v^2}.
\end{align}
\normalsize
The numerical solution of 
$G^R(k,\omega)^{-1}=0$ at $\epsilon_k=0$ yields $\omega=-U/2+0.87v,-U/2,-U/2-0.87v,U/2+0.87v,U/2,U/2-0.87v$. 
Hence, the upper (lower) Hubbard band at $\epsilon_k=0$ in the one-shot NCA calculations features a central peak at $U/2$ ($-U/2$) and two side peaks split off by an energy $\pm 0.87v$. This shows that even in the large-$U$ limit, the broadening of the momentum-resolved spectral function is comparable to the bandwidth of the noninteracting model. 
This analytical result is supported by the full NCA calculations for different $U$ in the Mott regime. In Fig.~\ref{fig:supp_fig6} we show that both the local spectral functions and the momentum-dependent spectral functions at $\epsilon_k=0$ are almost independent of the interaction strength.

\subsection{Other parameters}

\begin{figure}[t]
  \centering
    \hspace{-0.cm}
    \vspace{0.0cm}
   \includegraphics[width=60mm]{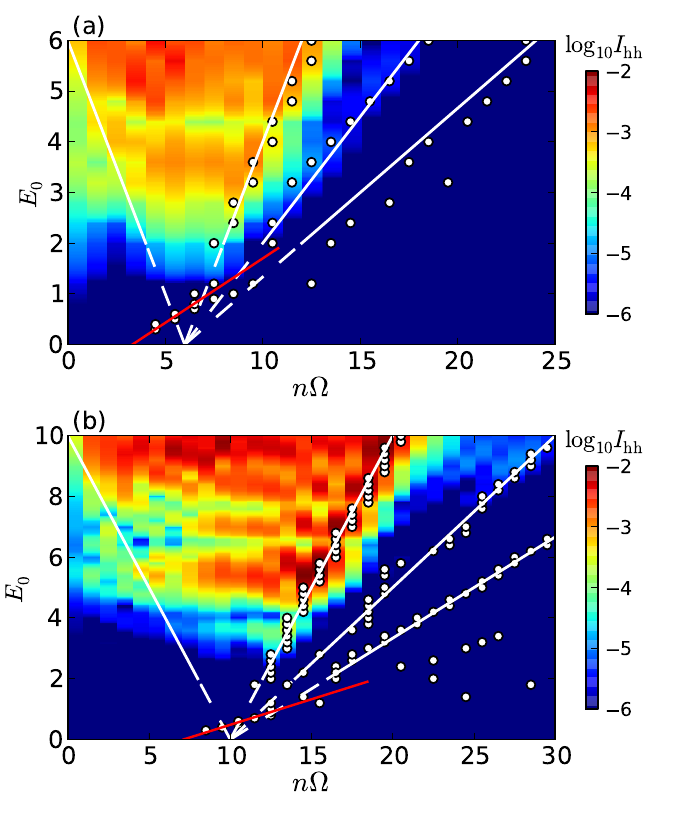} 
  \caption{HHG spectra ($\log_{10}(I_{\rm hh})$) in the plane of $E_0$ and $n\Omega$ for the Mott insulator on the hypercubic lattice with (a) $\Omega=0.5,U=6.0,\beta=2.0,\Gamma=0.06,W_{\rm bath}=5.0$, and (b) $\Omega=0.5,U=10.0,\beta=1.0,\Gamma=0.06,W_{\rm bath}=6.0$. White markers indicate the cutoff energies.}
  \label{fig:supp_fig3}
\end{figure}

In Fig.~\ref{fig:supp_fig3}, we plot the HHG spectrum for (a) $U=6$ and (b) $U=10$ in order to demonstrate that the HHG features 
discussed in the main text are generic. 
Namely, in the weak field regime, there emerges one plateau whose cutoff scales with $\epsilon_\text{cut}=\Delta_{\rm gap}+\alpha E_0$,
while in the strong field regime multiple plateaus appear, whose cutoffs scale with $\epsilon_{\text{cut},m}=U+mE_0$.
In addition, one can observe the characteristic features in the HHG intensity: i) the strong intensity regime in the triangular region [$U-E_0,U+E_0$],
ii) an enhanced intensity round $E_0=U/2$ and iii) a suppressed intensity around $U/2\lesssim E_0\lesssim U$.

\subsection{Hubbard 1 approximation and impurity effects on the semiconductor model}

\begin{figure}[t]
  \centering
    \hspace{-0.cm}
    \vspace{0.0cm}
   \includegraphics[width=86mm]{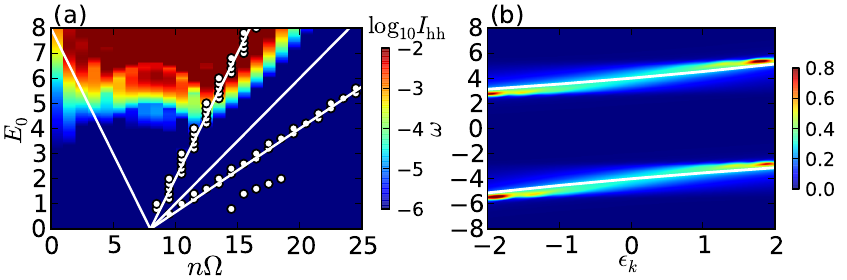} 
  \caption{(a) HHG spectra ($\log_{10}(I_{\rm hh})$) evaluated by Hubbard 1 in the plane of $E_0$ and $n\Omega$ for $\Omega=0.5$. (b) Momentum dependent spectral function $A(k,\omega)=[A_{cc}(k,\omega)+A_{vv}(k,\omega)]/2$ of the type 1 semiconductor with impurity scattering ($V_{\rm imp}=0.55$). Here $U=8.0,\beta=2.0,\Gamma=0.06$.
  }
  \label{fig:supp_fig7}
\end{figure}

\begin{figure}[b]
  \centering
    \hspace{-0.cm}
    \vspace{0.0cm}
   \includegraphics[width=86mm]{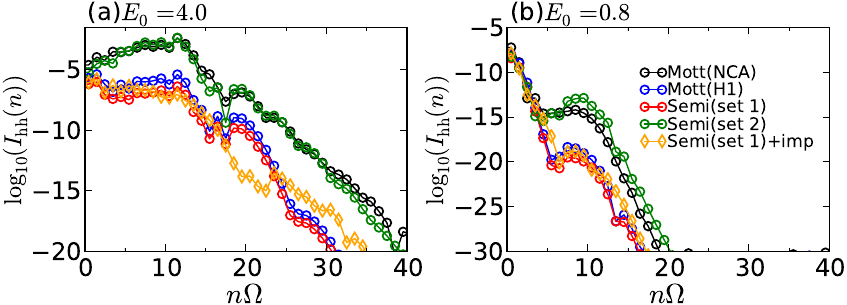} 
  \caption{(a)(b) Comparison of the HHG spectrum for $E_0=4.0$ (a) and $E_0=0.8$ (b) for NCA, Hubbard 1, the type 1 and type 2 semiconductors and the type 1 semiconductor with impurity with $V_{\rm imp}=0.55$. 
  Here  $\Omega=0.5,U=8.0,\beta=2.0,\Gamma=0.06$.}
  \label{fig:supp_fig8}
\end{figure}

Within the Hubbard 1 (H1) approximation, the self-energies coming from the interaction are expressed as 
\small
\begin{subequations}
\begin{align}
\Sigma^R(\omega)&=\frac{U^2}{4(\omega+i\eta)},\\
\Sigma^K(\omega)&=-i\frac{\eta U^2}{2(\omega^2+\eta^2)}\tanh \bigl(\tfrac{\beta\omega}{2}\bigl),
\end{align}
\end{subequations}
\normalsize
where $\eta=0^+$. The total self-energy is the sum of this and the contribution from the heat bath.
In Fig.~\ref{fig:supp_fig7}(a), we show the HHG spectrum evaluated with the H1 approximation in the plane of the field strength $E_0$ and the harmonic energy $n\Omega$.  
One can see that the global features of the result are very similar with the type 1 semiconductor (see Fig.~\ref{fig:fig3}(b) in the main text and also Fig.~\ref{fig:supp_fig8} for a more detailed comparison). 
Even though the system is a Mott insulator with a large gap, a naive usage of the H1 approximation leads to a qualitatively wrong HHG spectrum and underestimates the intensity in a wide parameter range.

In order to mimic the finite width in the single particle spectral function of the Mott insulator, we add the effect of impurity scattering in the semiconductor model 
through the self-energy,
\small
\begin{align}
\hat{\Sigma}(t,t')=V_{\rm imp}^2 \hat{G}_{\rm loc}(t,t').
\label{self_impurity}
\end{align}
\normalsize
Here the hat indicates a $2\times 2$ matrix in the band indices.
$V_{\rm imp}$ is evaluated from impurity averaging as $V_{\rm imp}^2=n_{\rm imp}\bar{V}^2_{\rm imp}$, where $\bar V_\text{imp}$ is the impurity potential and 
$n_{\rm imp}$ represents the density of impurities and we ignore the momentum dependence of the scattering matrix element~\cite{Kemper2013b}.
The total self-energy is the sum of Eq.~(\ref{self_impurity}) and the self-energy from the heat bath.
In Fig.~4(b) in the main text and Fig.~\ref{fig:supp_fig7}(b), we show the resulting single particle spectral function for $V_{\rm imp}=0.55$, which 
well reproduces the spectral features of the Mott insulator, see Fig.~4 (a) in the main text.

In Fig.~\ref{fig:supp_fig8}, we compare the HHG spectra for the Mott insulator evaluated with NCA and H1 and those of the type 1 and type 2 semiconductors for $E_0=4.0,0.8$ and $\Omega=0.5$.
The H1 result matches that of the type 1 semiconductor, while the NCA result is very similar to that of the type 2 semiconductor. 
The impurity scattering affects the HHG spectra at high frequencies but in the plateau regions it has rather small effects.

\end{document}